\documentclass[pdflatex,sn-mathphys-num]{sn-jnl}% Math and Physical Sciences Numbered Reference Style
\usepackage{tiaStyleFile}
\usepackage{gensymb}
\usepackage{amssymb}
\usepackage{graphicx}%
\usepackage{multirow}%
\usepackage{amsmath,amssymb,amsfonts}%
\usepackage{amsthm}%
\usepackage{mathrsfs}%
\usepackage[title]{appendix}%
\usepackage{xcolor}%
\usepackage{textcomp}%
\usepackage{manyfoot}%
\usepackage{booktabs}%
\usepackage{algorithm}%
\usepackage{algorithmicx}%
\usepackage{algpseudocode}%
\usepackage{listings}%
\usepackage{amsmath, amsfonts} % basic math packages
\usepackage{amsthm}
\usepackage{mathtools}
\usepackage{amsfonts} % basic math packages
\usepackage{tikz} % for making illustrations
\usetikzlibrary{shapes.arrows, arrows, decorations.markings, decorations.text}
\usetikzlibrary{calc}
\usepackage{graphicx} % for importing images
\usepackage{xcolor} % more color options
\usepackage{multicol} % for making two-column lists
\usepackage{fullpage}
\usepackage{hyperref} % for hyperlinking
\usepackage{orcidlink}
\usepackage{algorithm}
\usepackage{algpseudocode}
\usepackage{cleveref}
\usepackage{siunitx}

\theoremstyle{thmstyleone}%
\theoremstyle{thmstyletwo}%

\theoremstyle{thmstylethree}%

\begin{document}

\title[Article Title]{VPAL: A novel method to reduce reconstruction time for 5D free-running imaging}


\author*[1]{\fnm{Yitong} \sur{Yang}}

\author[2]{\fnm{Muhammad} \sur{Naeem}}

\author[2]{\fnm{Marly} \sur{van Assen}}

\author[3,5]{\fnm{Jérôme} \sur{Yerly}}

\author[4,5]{\fnm{Davide} \sur{Piccini}}

\author[3,5]{\fnm{Matthias} \sur{Stuber}}

\author[1,2]{\fnm{John} \sur{Oshinski}}

\author[6]{\fnm{Matthias} \sur{Chung}}

\affil*[1]{\orgdiv{Wallace H Coulter Department of Biomedical Engineering}, \orgname{Georgia Institute of Technology \& Emory University School of Medicine}, \orgaddress{ \city{Atlanta},  \state{GA}}}

\affil[2]{\orgdiv{Department of Radiology and Imaging Sciences}, \orgname{Emory University School of Medicine}, \orgaddress{ \city{Atlanta},  \state{GA}}}

\affil[3]{\orgdiv{Department of Diagnostic and Interventional Radiology}, \orgname{Lausane University Hospital (CHUV) and University of Lausane (UNIL)}, \orgaddress{ \city{Lausanne},  \state{Vaud}, \country{Switzerland}}}

\affil[4]{\orgdiv{Advanced Clincal Imging Technology}, \orgname{Siemens Healthineers International AG}, \orgaddress{ \city{Lausanne},  \state{Vaud}, \country{Switzerland}}}

\affil[5]{\orgname{Center for Biomedical Imaging (CIBM)}, \orgaddress{ \city{Lausanne}, \state{Vaud}, \country{Switzerland}}}

\affil[6]{\orgdiv{Department of Mathematics}, \orgname{Emory University}, \orgaddress{ \city{Atlanta}, \state{GA}}}


\abstract{
\textbf{Purpose}: 
Ferumoxytal-enhanced 5D free-running whole heart cardiovascular magnetic resonance imaging (CMR) provides image quality comparable to computed tomography angiography (CTA).  However, this technique requires hours-long reconstruction time, preventing widespread clinical usage.  Therefore, a more efficient reconstruction algorithm is needed. \\ 
\textbf{Approach}: 
A variable projection augmented Lagrangian (VPAL) method for 5D motion-resolved image reconstruction was developed and compared with the state-of-the-art alternating direction method of multipliers (ADMM) in five numerical simulations and 15 in-vivo pediatric data set. The relative error of the reconstructed images using the VPAL technique versus the ground-truth images were compared with those reconstructed using ADMM. In in-vivo subjects, reconstruction time, mid-short axis (SA) blood-myocardium sharpness, left ventricular ejection fraction (LVEF), and a radiologist's image quality ratings (0-4 scale) were compared in VPAL and ADMM reconstructions. A paired t-test was performed for significance testing ($p<0.05$). Linear regression analysis and Bland-Altman analysis were used to assess the agreement between the ADMM and VPAL reconstructions. \\
\textbf{Results}: 
In numerical simulation dataset, we found that the relative error of VPAL reconstructions with the ground truth was not significantly different from the relative error of ADMM reconstructions with the ground truth, $p = 0.07$. In in-vivo datasets, VPAL reduced the reconstruction time from 16.3 $\pm$ 3.6 hours (ADMM) to 4.7 $\pm$ 1.1 hours (VPAL), $p=1e-10$. Blood-myocardium border sharpness in VPAL reconstructions closely correlates with ADMM reconstructions, $R^2 = 0.97$. The LVEFs measured using the VPAL reconstructions are largely consistent with those measured using the ADMM reconstructions, 56 $\pm$ 6 \% in ADMM and 56 $\pm$ 6 \% in VPAL, $p=0.55$. Both VPAL and ADMM have good to excellent diagnostic ratings (VPAL vs. ADMM: 3.9 $\pm$ 0.3 vs. 3.8 $\pm$ 0.4 in 2-chamber; 3.9 $\pm$ 0.4 vs. 3.9 $\pm$ in 4-chamber; 3.7 $\pm$ 0.5 vs. 3.7 $\pm$ 0.5 in mid-SA reformatted views. \\
\textbf{Conclusion}: 
VPAL is a more time efficient algorithm for 5D motion-resolved iterative reconstruction compared to the current state-of-the-art ADMM methods, as evaluated by mid-SA sharpness of the blood myocardium, 3D SSIM, image quality ratings from the radiologists, and functional quantification of LVEF, supporting its potential for clinical use.
}




\maketitle

\section{Introduction}\label{sec1}

\textbf{C}ardiac \textbf{m}agnetic \textbf{r}esonance (CMR) imaging acquires 3D motion-resolved images of the myocardium over the cardiac cycle and is currently the gold standard for evaluating cardiac function and wall motion abnormalities \cite{bib1}. Obtaining high-resolution CMR images typically requires breath holding (BH) to suppress image blurring from respiration motion. However, BH is challenging for many subjects and is generally only applicable on 2D slice acquisitions, which requires time-consuming  planning from a trained operator. Hence, to acquire 3D motion-resolved images, respiratory navigators are often used to gate acquisition during end-expiration.  Respiratory navigators are problematic with irregular breathing patterns and their use leads to unpredictable scan times. In addition, an electrocardiogram (ECG) signal is currently used to gate acquisition to enable acquisition of multiple cardiac phases, and each phase is partially filled during each cardiac cycle. However, the ECG signal is prone to the magnetohydrodynamic effect in MRI, which often leads to a falsely detected cardiac rhythm \cite{bib2}. Recently, an image-based self-gating methods \cite{bib3, bib4} have been used to substitute ECG gating to obtain cardiac rhythm from free-running raw data. Similarly, respiratory motion is also extracted from the raw data, eliminating the need for external respiratory navigators\cite{bib5}.  Self-gating methods yield a fixed scan time but suffer from long reconstruction times. 

5D free-running imaging is an ungated free-breathing technique that provides 3D images of the whole heart, resolved in cardiac and respiratory dimensions and with isotropic resolution \cite{bib6}. ``5D'' refers to the 3 spatial dimensions (x-y-z), and the additional two dimensions that correspond to the cardiac and respiratory cycles, respectively. The k-space data are incoherently sampled in a golden-angle radial phyllotaxis trajectory with non-overlapping and non-uniform distribution\cite{bib7}. After motion extraction and data binning, the under-sampled k-space data from the 5D free-running motion-resolved acquisition are reconstructed into a 5D image using compressed sensing-based iterative optimization \cite{bib8, bib9,bib10}, a very time-consuming step. 

Reconstruction of MR images from the Fourier space (k-space), often is an ill-posed problem  since either fewer data are acquired compared to the dimensionality of the image \cite{bib11} due to acquisition constraints, or the forward operator $A$ is not invertible in a stable way, such as in 5D free-running imaging where multi-coil sensitivity encoding (SENSE) \cite{bib10,bib12} and non-cartesian sampling are used. A $\ell_1$, total variation (TV)-regularized compressed sensing reconstruction is used to iteratively solve for the image from the k-space. A regularization parameter is used to scale regularization term to balance the sparsity, or smoothness, with the noise level of the image \cite{bib13}. Since this iterative reconstruction is time-consuming, the current interest has been in the use of machine learning (ML) and deep learning (DL) strategies to replace the iterative reconstruction process or to develop efficient numerical optimization algorithms with faster convergence rates \cite{bib14,bib15}. ML and DL methods are powerful techniques, but are highly dependent on the quantity and quality of the data used for model training, and often application-dependent\cite{bib15}. Instead, an efficient numerical algorithm can be implemented for any image resolution, dimension, and contrast reconstruction problem. 

TV regularization was designed to recover sharp edges from blurred and corrupted images. The numerical methods for solving the TV-regularized problem include primal-dual methods, the Newton method, augmented Lagrangian approaches, \cite{bib16}. Among these methods, the augmented Lagrangian approach is the most commonly used, as it does not require an approximation of the $\ell_1$-norm term's gradient, but reduces the problem into two subproblems (details in \cref{sec:theory}). The augmented Lagrangian method has widely demonstrated its usability in compressed sensing applications \cite{bib17}. The $\ell_1$-norm of TV regularization imposes sparsity on the finite difference  of the image and has shown highly accurate reconstruction images even from highly undersampled k-space data \cite{bib6,bib9}. The alternating direction method of multiplier (ADMM) is among the augmented Lagrangian-based methods and is one of the most common methods to solve such $\ell_1$, TV regularized problem. ADMM splits the non-differentiable problem into a linear least squares problem and a soft thresholding problem that has a closed-form solution\cite{bib18}. However, as image dimension and matrix size increase, ADMM becomes less efficient and can be very costly to compute. Recently, the variable projection \cite{bib19} method for solving the $\ell_1$ augmented Lagrangian problem was established and has demonstrated improved performance in terms of convergence compared to alternating direction methods such as ADMM. 

In this study, we proposed a novel variable projection-augmented Lagrangian (VPAL) method by introducing a targeted update strategy to reconstruct the 5D free-running data off-line on MATLAB from the raw data and evaluate its computational effectiveness (time and accuracy) compared to ADMM. The performance of VPAL will be compared with that of the state-of-the-art ADMM method in numerical simulations by measuring the convergence time and relative error.  Reconstruction time, image quality, and functional evaluations (that is, left ventricular ejection fraction (LVEF)) will also be compared on the in-vivo pediatric congenital heart disease patients' dataset. The sharpness of the blood-myocardium in the mid-short axis (SA) view, the 3D structural similarity index measure (3D SSIM), and a radiologist's image quality rating will be used as image quality metrics to compare VPAL and ADMM reconstructions. We\textbf{ hypothesized }that using the same number of iterations as ADMM, VPAL can finish the reconstruction more quickly, while the image quality metrics and functional scores are similar to ADMM. 

This manuscript is ordered as follows. The mathematical setup of the 5D free-running image reconstruction problem and the current state-of-the-art algorithm is presented in \cref{sec:theory}. The proposed iterative reconstruction algorithm for VPAL is presented in \cref{vpal+ alg}. The study data contains numerical simulation data \cref{numerical phantom} and in-vivo dataset \cref{in-vivo}. The performance of VPAL and ADMM reconstructions on numerical simulation is in \cref{subsubsec2}, and in-vivo data is illustrated in \cref{subsec2}. 

\section{Theory}\label{sec:theory}
Let the vector $x \in \mathbb C^{n}$ represent the time-varying cardiac volume, $A \in \mathbb R^{m\times n}$ describing the mathematical MRI acquisition process mapping $x$ to the k-space data $b \in \mathbb C^m$. We assume the acquisition processes corrupted with additive noise $\epsilon \in \mathbb C^m$ be some additive noise originating from measurements errors in the recording and potential model errors,
  $$ A x + \epsilon= b.$$
In the 5D free-running MRI imaging, the k-space data is acquired with non-Cartesian sampling \cite{bib7} and sensitivity encoding, such that the encoding operator $A$ involves non-uniform fast Fourier transform (NUFFT) operations \cite{bib20} modulated by the corresponding coil's sensitivity profile \cite{bib12}.
The inverse problem can now be stated as given the forward process $A$ and noisy k-space data $b$ reconstruct $x$. Since  the forward operator $A$ is not exactly invertible in a stable way, the reconstruction problem is ill-posed, meaning a solution does not exist, is not unique, or does not depend continuously on the data, see \cite{bib21}. Prior knowledge is required to stabilize the solution process. We followed the common variational inverse problem formulation \cite{bib22}, where the inverse problem is stated as
\begin{equation} \label{eq:opt}
  \min_x \  f(x) = \tfrac{1}{2} \|Ax-b\|_2^2 + R(x),
\end{equation}
where $\tfrac{1}{2} \|Ax-b\|_2^2$ is the \emph{data fidelity} term measuring the data misfit, and $R(x)$ is a \emph{regularization} term encoding prior knowledge of the solution.

In 5D free-running imaging in which both cardiac and respiratory motions are reconstructed, it is crucial to compensate for motion artifacts and encourage smoothness over time. Temporal total variations are applied to encourage smoothness over time and reduce motion artifacts (Tikhonov \cite{bib23}, lasso \cite{bib24}, grouplasso \cite{bib25}, etc.).The standard approach for this reconstruction problem employs $\ell1$ norm regularization on the temporal finite difference, see \cite{bib6}, which promotes sparsity and piecewise-constant behavior. In this study, we instead consider $\ell2$ norm regularization on the temporal finite difference, which encourages smooth temporal variations.  Hence, we imposed a regularizer that separately promotes temporal continuity along the respiratory and cardiac time frames and sparsity in the spatial domain of the image supporting sharp edges in the reconstructed image, i.e.,
  $$ R(x) =  \lambda_s \|D_s x\|_1 + \lambda_c \|D_c x\|_2^2 + \lambda_r \|D_r x \|_2^2 , $$
  where $D_c$, $D_r$, and $D_s$ represent the finite difference approximations of cardiac, respiratory and spatial dimensions and $\lambda_c, \lambda_r, \lambda_s>0$ their corresponding regularization parameter. 

Iterative approaches to solve \Cref{eq:opt} are required; however, a major challenge in 5D free-running MRI reconstructions is the dimensionality and ill-posedness of the underlying problem. Another challenge appears through the nondifferentiable $L_1$ regularization term. Various numerical approaches have been proposed to solve \Cref{eq:opt}, including alternating direction, iterative shrinkage-thresholding methods, and variable projection to name a few \cite{bib18,bib19,bib26}. Each approach has advantages and disadvantages. Here we followed a splitting approach which introduces an auxillary variable to circumvent the non-differentiability. Let $y = D_s x$, then \Cref{eq:opt} can be restated as a constrained optimization problem
  \begin{equation}\label{eq:constr}
  \min_{x,y} \ \tfrac{1}{2} \|Ax-b\|^2_2  + \tfrac{\lambda_c}{2} ||{ D_c{x}}^2 + \tfrac{\lambda_r}{2}||{D_{r}x}||^2 + \lambda_s \| y \|_1 \qquad \text{subject to   } \quad
  y - D_s x = 0.
\end{equation}
One approach to solve \Cref{eq:constr} utilized the method augmented Lagrangian, i.e.,
  $$\min_{x,y,\mu} \ \tfrac{1}{2} \|Ax-b\|^2_2  + \tfrac{\lambda_c}{2} ||{ D_c x}||^2 + \tfrac{\lambda_r}{2}||{D_{r} x}||^2 +\lambda_s \| y \|_1 + \frac{\rho}{2} \| D_s x - y + \mu \|_2^2  - \tfrac{\rho}{2} \|\mu\|_2^2,$$
  where $\mu$ denotes the Lagrange multipliers and $\rho>0$ the augmented Lagrangian penalty parameter, details see \cite{bib27}. A common approach is to employ an alternating direction method of multipliers (ADMM), \cite{bib18}. ADMM, as the name suggests, iteratively alternates the optimization of variables $x$, $y$, and $\mu$. Given $y^{k-1}$ and $\mu^{k-1}$ at iteration $k$ we update
    \begin{align}
    x^{k}  &= \argmin_x \  \tfrac{1}{2} \|Ax-b\|_2^2 +  \tfrac{\lambda_c}{2} \norm[2]{D_c x}^2 + \tfrac{\lambda_r}{2}\norm[2]{D_{r}x}^2+ \tfrac{\rho}{2}\|D_s x - y^{k-1} + \mu^{k-1}\|_2^2, \label{eq:admm1}\\
     y^{k} &=  \argmin_y  \ \lambda_s \norm[1]{y} + \tfrac{\rho}{2} \norm[2]{y - D_s x^{k} + \mu^{k-1} }^2, \\
    \mu^{k} &= \mu^{k-1} + y^{k} - D_s x^{k}.
\end{align}
Optimization for the auxiliary variable $y$ is the well-known shrinkage problem and is computationally efficient. Its closed-form solution is given by
  \begin{equation}\label{eq:shrink}
      y^{k}= \sign{D_{s} x^{k}+\mu^{k-1}}\hadamard \left(\left|\nabla_{s} x^{k}+\mu^{k-1} \right| - \tfrac{\lambda_s}{\rho}\right)_+,
  \end{equation} 
where $\hadamard$ denotes the Hadamard product, and $(\mdot)_+$ is the element-wise function where each element is given as 
$$
(w)_+ = 
\begin{cases}
    w, & \text{if } >0,\\
    0, & \text{otherwise.}
\end{cases}
$$
In machine learning, this function is also referred to as the Rectified Linear Unit (ReLU).

The splitting results of the problem \Cref{eq:admm1} is a linear least-squares optimization and can be solved analytically. However, for the dimension at hand, direct solvers are infeasible. Consequently, linear iterative approaches, such as LSQR \cite{bib28} are required to solve for $x^{k+1}$ in each iteration $k$. Hence, solving the subproblem \Cref{eq:admm1} remains computationally expensive as multiple inner conjugate gradient iterations are needed and each requires gradient computation on large-scale datasets as in 5D image reconstruction.

\section{Method}\label{sec2}

\subsection{Reconstruction framework}

\subsubsection{VPAL reconstruction algorithm}\label{vpal+ alg}
Instead of ADMM, we propose to use a variable projection method to solve the augmented Lagrangian problem, named the variable projection augmented Lagrangian (VPAL) method \cite{bib19}. Instead of solving explicitly at each step, $x^{k+1}$ is computed by updating $x^{k}$ along a descent direction $d^k$ scaled by a step size $\alpha^k$. The VPAL algorithm updates the image $x^k$, the auxiliary variable $y^k$, and the Lagrange multiplier $\mu^k$ as follows
\begin{align*}
    & x^{k} = x^{k-1} + \alpha^k d^{k-1}, \\
    & y^{k} =  \argmin_y \ \lambda_1 \|y\|_1 + \tfrac{\rho}{2} \norm[2]{D_s x^{k} - y + \mu^{k-1} }^2, \\
    & \mu^{k} = \mu^{k-1} + y^{k} - D_s x^{k},
\end{align*}
where $d^k$ is an appropriate descent of the function
\begin{equation}\label{eq:f(x)}
     f(x) = \ \thf \norm[2]{Ax-b}^2 + \lambda_1 \norm[1]{\widetilde y(x)} + \tfrac{\lambda_2}{2} \norm[2]{D_c x}^2 + \tfrac{\lambda_3}{2}\norm[2]{D_r x}^2 + \tfrac{\rho}{2}\norm[2]{D_s x - \widetilde y(x) + \mu^k}^2 
\end{equation}

at $x_k$ with 

\begin{equation}\label{eq:shrinkage}
    \widetilde y(x) = \sign{D_{s} x+\mu^{k}}\hadamard \left(\left|D_s x +\mu^{k} \right| - \tfrac{\lambda_1}{\rho}\right)_+.
\end{equation}

Note that our approach is not the same as solving the ADMM problem with a single conjugate gradient step. Our method computes the update of $x^{k}$ by projecting the variable $y$ to a reduced subspace of $x$, through a soft-thresholding step. Specifically, a projected variable $\widetilde y(x)$ is taken into account in the calculation of the descent direction and the step size with respect to $x$. 

A variable projection method has been proposed for $\ell1$ regularization methods in linear systems using a simplified gradient descent and linearized step size selection for illustration \cite{bib19}. In multicoil, non-cartesian 3D radial imaging reconstruction problem, the forward operator $A$ involves coil sensitivity encoding and non-uniform Fourier transform (NUFFT), making it non-linear.  We thereby select a descent direction based on nonlinear conjugate gradient approaches, see \cite{bib27} for details. Specifically, we use the Fletcher-Reeves method to compute the nonlinear conjugate gradient descent step 
\begin{equation}\label{eq:fr-cg}
    d^k = -r^k + \beta^k d^{k-1}
\end{equation}
where $r^k = [\nabla_x f(x)]_{x = x^{k-1}}$ and the parameter $\beta^k$ within the nonlinear Fletcher-Reeves update is computed using the current gradient $r^k$ and gradient of the previous step $r^{k-1}$, i.e.,
\begin{equation}\label{eq:beta}
    \beta^k = \frac{(r^k)\t r^k}{(r^{k-1})\t r^{k-1}}.
\end{equation}

The iterative reconstruction using VPAL with a non-linear conjugate gradient step calculated based on the Fletcher-Reeves method is presented in \Cref{alg:vpal}. 

\vspace{-10pt}
\begin{algorithm}[H]
\caption{Variable Projected Augmented Lagrangian (VPAL)}\label{alg:vpal}
\begin{algorithmic}[1]
\Require $k = 0,x^0 = A^{'} b, \quad z^0 = \mu^0 = \text{shape}(D_s x^0), r^0 = \nabla f(x^0), d^0 = -r^0$
    \While{ not converged}
        \State $k = k+1$ \Comment{update iteration counter}
        \State $\alpha^{k} = \argmin_{\alpha} f(x^{k-1} + \alpha d^{k-1})$ \Comment{Line search to compute for step size}
        \State $x^{k} = x^{k-1} + \alpha^{k}d^{k-1}$\Comment{update image}
        \State $r^k = \nabla f(x^{k})$ \Comment{update new gradient step}
        \State $\beta^{k} = \frac{(r^k)\t r^k}{(r^{k-1})\t r^{k-1}}$ \Comment{update Fletcher-Reeves nonlinear CG parameter}
        \State $d^k = -r^k + \beta^{k} d^{k-1}$ \Comment{update descent direction}
        \State $y^{k}$ computed using \Cref{eq:shrink} 
        % $y^k = \text{shrinkage}(D_s x^{k+1} + \mu^k, \frac{2\lambda_1}{\rho})$ 
        \Comment{update auxiliary variable}
        \State $\mu^{k} = \mu^{k-1} + D_s x^{k
        } - y^ {k} $ \Comment{update Lagrange multiplier}
    \EndWhile
\end{algorithmic}
\end{algorithm}
\vspace{-10pt}

Note: $'$ is the adjoint operation
\subsection{Study data}

\subsubsection{Numerical simulation data}\label{numerical phantom}
To evaluate the performance of the proposed VPAL and the current state-of-the-art ADMM method, we developed reference 5D dynamic numerical simulations based on 4D functional CT angiography (CTA) data. 4D functional CTA dataset were obtained under IRB approval at Emory University from five clinical cardiac patients. All 4D CTA images contain 10 cardiac slices and each slice was cropped and interpolated to have a 1$\text{mm}^3$ isotropic resolution and a field of view (FOV) of 192x192x192.  The heart and diaphragm were segmented in each cardiac frame using 3D slicer software \cite{bib29}. The heart and diaphragm in each cardiac slice were displaced in the superior-inferior and anterior-posterior directions based on four equally spaced positions sampled from a sinusoidal wave with a frequency of 18 beats per minute and an amplitude of 8 mm. The displacement at each position was determined by the amplitude of the wave at that location. The four distinct respiratory locations correspond to end-expiration, mid-expiration, mid-inspiration, and end-inspiration in the 5D dynamic ground-truth numerical simulations. The 5D dynamic cardiac MR ground truth images are then under-sampled in k space using the golden-angle radial phyllotaxis trajectory \cite{bib7} with 12 segments per interleave and a variable number of interleaves, giving an under-sampling ratio of 20\% in each 3D image of the physiological phases of cardiac and respiratory\cite{bib5}. The under-sampling ratio is calculated by $\frac{2 N_{\text{seg}} N_{\text{shot}}}{\pi N_p^2}$, where $N_{\text{seg}}$ is the number of segments, $N_{\text{shot}}$. The retrospective k-space sampling assumes a single-coil acquisition scenario. The under-sampled k-space data of 10 cardiac slices and 4 respiratory frames and the image of the initial guess of which re-gridding and direct NUFFT inverts are inputs to the reconstruction algorithms.

\subsubsection{In-vivo data} \label{in-vivo}
This study utilized raw ungated 5D free-running whole-heart CMR data from 15 pediatric congenital heart disease patients (age 16 $\pm$ 4, 10 female, heart rate 72 $\pm$ 18, and respiratory rate 16 $\pm$ 4.2).  All images were acquired after a slow IV infusion of ferumoxytol (Ferheme, AMAG pharmaceutical) at a concentration of 4mg/kg. The image acquisitions were approved by the Institutional Review Board of Lausanne University Hospital and written informed consent was obtained from the legal guardian of each subject prior to CMR scanning. The images were acquired on a 1.5 Tesla scanner (MAGNETOM Sola, Siemens Healthcare, Erlangen, Germany), using an 18-channel body coil placed over the heart. Data were acquired with a previously described\cite{bib6}, prototype, continuous, free-running (ECG-free, navigator-free) golden-angle radial spiral phyllotaxis \cite{bib7} GRE sequence over a 220 x 220 x 220 $mm^3$ field-of-view (FOV) with 192 samples per readout, giving an isotropic spatial resolution of 1.1 $mm^3$. RF excitation used a 15\degree flip angle. All fat saturation pulses and ramp-up RF excitations were removed. A total of 124,344 radial lines were acquired with 12 segments (radial lines) per interleaf. The total acquisition time was 5 minutes and 54 seconds. 

The cardiac and respiratory motion signals were extracted from the raw data and used to bin the raw data into cardiac and respiratory frames using a previously reported methodology using MATLAB (MathWorks)\cite{bib6}. A brief summary of this methodology is presented here. Principal component analysis of the temporally resolved superior-inferior projections of each interleaf of data produced a physiological motion signal that spanned the duration of the sequence acquisition and represented a superposition of the cardiac and respiratory motion signals. These two signals were separated through a power spectrum density analysis. Once the signals were successfully identified and separated, data were binned into four respiratory and a variable number of cardiac phases determined such that each phase represented a 50 ms window without view sharing, consistent with the previously published 5D free-running approach \cite{bib6}. After motion extraction and binning, the raw k-space binned data was reconstructed by optimization algorithms VPAL \cref{alg:vpal} and ADMM \cref{sec:theory}.

\subsection{Study design}
\subsubsection{Numerical simulation study}

The cardiac and respiratory resolved k-space data of the numerical simulations are reconstructed into 5D images using the proposed VPAL algorithm and the current standard ADMM algorithm. The stopping criterion for ADMM and VPAL iterative reconstruction is set such that the iterations stop when the change in relative error $\tfrac{\|\text{ground-truth image} - \text{reconstructed image}\|}{\|\text{ground-truth image}\|}$ between successive iterations is less than 0.001. The VPAL and ADMM consist of varying numbers of iterations, but the inner iterations of ADMM reconstruction that minimize the convex quadratic function in \Cref{eq:admm1} are kept at 4 conjugate gradient iterations. The proposed VPAL reconstruction also consists of varying numbers of iterations based on the stopping criterion. 
The following parameters $\lambda_1 = 0.0001$, $\lambda_2=\lambda_3=0.5$, and $\rho = 0.06$ appearing in \cref{vpal+ alg} were empirically determined \cite{bib6} and used for both reconstructions. The time to reach the stopping criterion of each algorithm is defined as the convergence time and is compared to the VPAL and ADMM methods. The relative error is computed as the Euclidean distance between the reconstructed image and the ground-truth image weighted by the $\ell2$ norm of the ground-truth image. 
\subsubsection{In-vivo study}

The binned k-space data are reconstructed into a 5D image using the proposed VPAL algorithm and the current standard ADMM algorithm. Due to the lack of ground truth images in the in-vivo cohort, both ADMM and the proposed VPAL reconstructions used the same number of outer iterations that were previously published and optimized for ADMM \cite{bib6}. Similarly to numerical simulations, 4 conjugate gradient iterations minimize the convex quadratic function in \Cref{eq:admm1}. The following parameters $\lambda_1 = 0.0001$, $\lambda_2=\lambda_3=0.5$ and $\rho = 0.06$ were empirically determined and used for both reconstructions. 

To compare the performance of VPAL versus ADMM reconstructions, the 3D structural similarity index measure (SSIM) is calculated based on the local means, standard deviations and cross-covariance of the reference and target images \cite{bib30}. 3D SSIM of VPAL is measured in the cardiac resting and peak systolic phases using ADMM as a reference.  The sharpness of the blood-myocardium border on the images of the mid-short axis is measured at the end-expiration, cardiac resting phase of the ADMM and VPAL reconstructed images using a previously presented method \cite{bib31}.  The left ventricular ejection fraction (LVEF) is evaluated on the reformatted short-axis stacks of the VPAL and ADMM reconstructed 5D images on Segment (Medviso). A cardiac radiologist with more than 20 years of experience rated the diagnostic quality of the ADMM and VPAL, reconstructed images on a 0-4 scoring scale given 2-chamber, 4-chamber and mid-short-axis cine movies where 0 is nondiagnostic, 1 is marked blurring, 2 is moderate blurring but has diagnostic value, 3 is mild blurring but good diagnostic value, and 4 is excellent diagnostic value.  

\subsection{Statistics}
The student-paired t-test with alpha = 0.95 is used for the significance testing of reconstruction time, relative error, SSIM, sharpness, and image quality readings of the VPAL and ADMM reconstructed images. 
Linear regression analysis assesses the similarity of mid-SA sharpness values computed on ADMM and VPAL reconstructions. Specifically, a linear model of the form  $y = ax + b$  is fitted, where $y$ represents the mid-SA sharpness values of VPAL and $x$ is from ADMM. A slope ($a$) close to 1 and an intercept ($b$) close to 0 indicate a strong similarity. The coefficient of determination ( $R^2$ ) is computed to assess how well the regression line explained the variability in the sharpness values. 
Bland-Altman analysis is performed to evaluate the agreement of the left ventricular ejection fraction (LVEF) between VPAL and ADMM. The mean difference (bias) and the 95\% limits of agreement (mean difference ± 1.96 × standard deviation of differences) were calculated to assess systematic bias and variability between methods.

\section{Results}\label{sec3}

\subsection{Numerical simulation reconstructions}\label{subsubsec2}
An example of comparisons of numerical simulation reconstruction with ground truth is shown in Figure 1. 
\begin{figure}[t]
\centerline{\includegraphics[width=20pc,height=20pc]{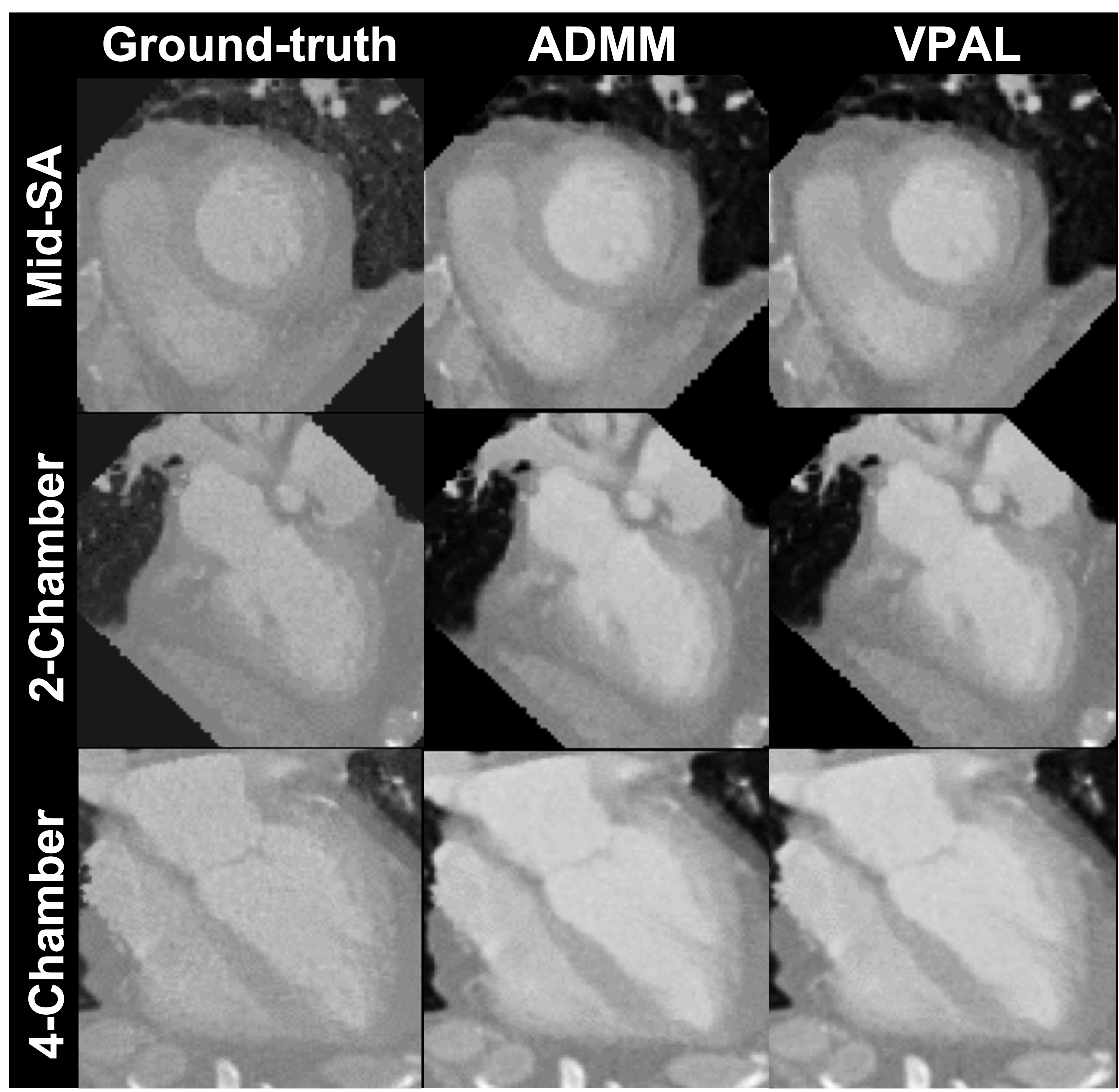}}
\caption{\textbf{Example numerical simulation reconstructions} The 4-chamber, mid-short-axis, and 2-chamber view in cardiac resting phase demonstrated in the ground truth 5D image, and the output of the reconstructions using ADMM and VPAL algorithms. }\label{fig1}
\end{figure}
The number of iterations to reach convergence is on average 21 $\pm$ 9 using ADMM and 38 $\pm$ 11 using VPAL reconstructions. The computational time to reach convergence is on average 27\% shorter using the VPAL algorithm (3.9 $\pm$ 1.2 hours) compared to using the ADMM algorithm (5.3 $\pm$ 2.4 hours); however, the difference is not significant (p = 0.11). The box and whisker plot of the convergence time of the ADMM and VPAL reconstructions is plotted in Figure 2. 
\begin{figure}[t]
\centerline{\includegraphics[width=20pc,height=15pc]{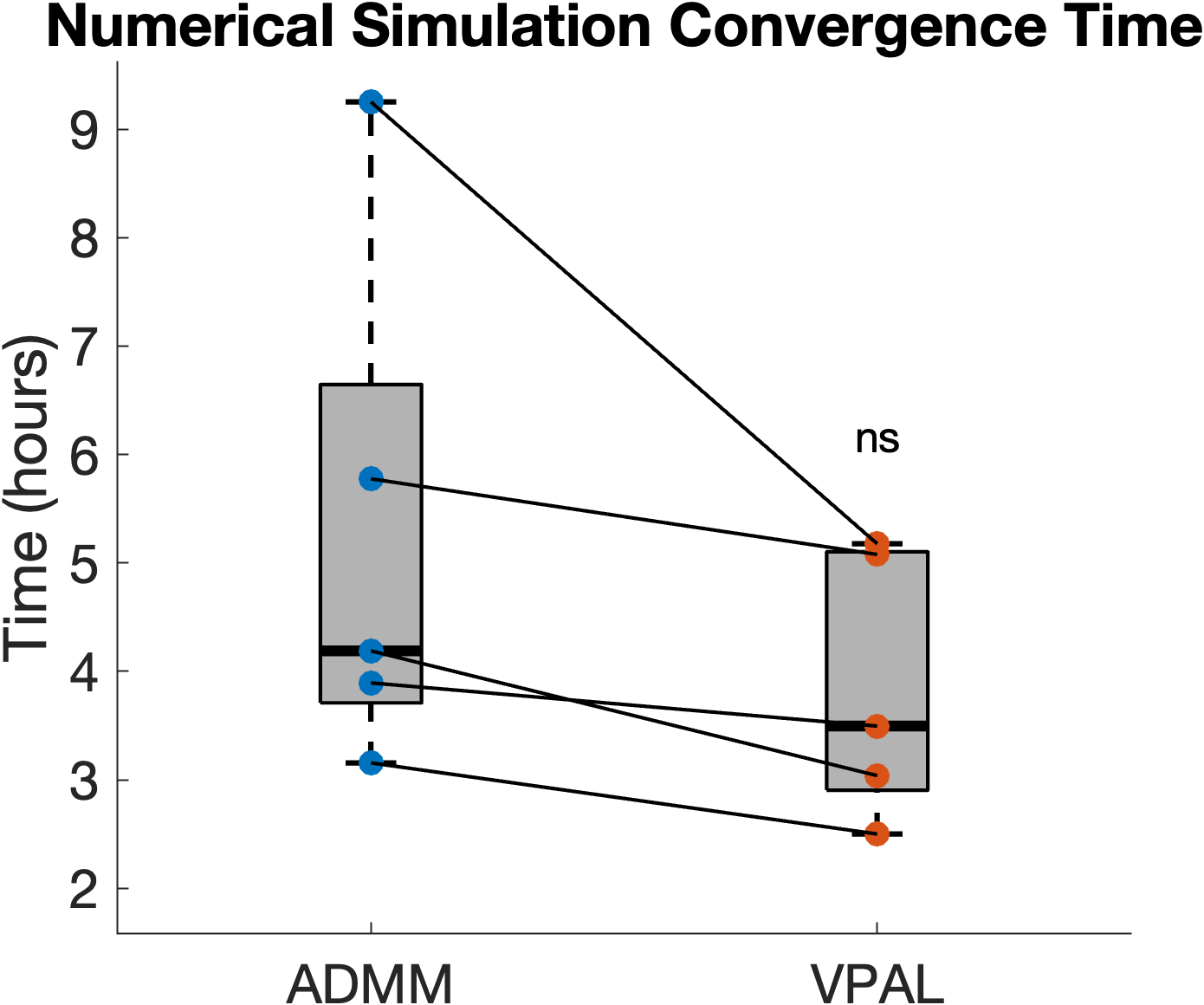}}
\caption{\textbf{Boxplot of convergence time using ADMM and VPAL reconstruction algorithms.}No statistical significance is found on the convergence time using ADMM versus VPAL}\label{fig1}
\end{figure}
The relative error of the reconstructed images with the ground-truth images in the numerical simulation is highly similar, 0.10 $\pm$ 0.03 in ADMM versus 0.10 $\pm$ 0.03 in VPAL ($p=0.07$). A strong similarity is found between the relative error of ADMM and VPAL reconstructions, the linear fitting has a slope of 1.03 and an intercept of $-0.00$ ($R^2$ = 0.9990). Figure 3 shows this linear relationship and the fitted line overlapped with the diagonal line $y = x$. 
\begin{figure}[t]
\centerline{\includegraphics[width=20pc,height=15pc]{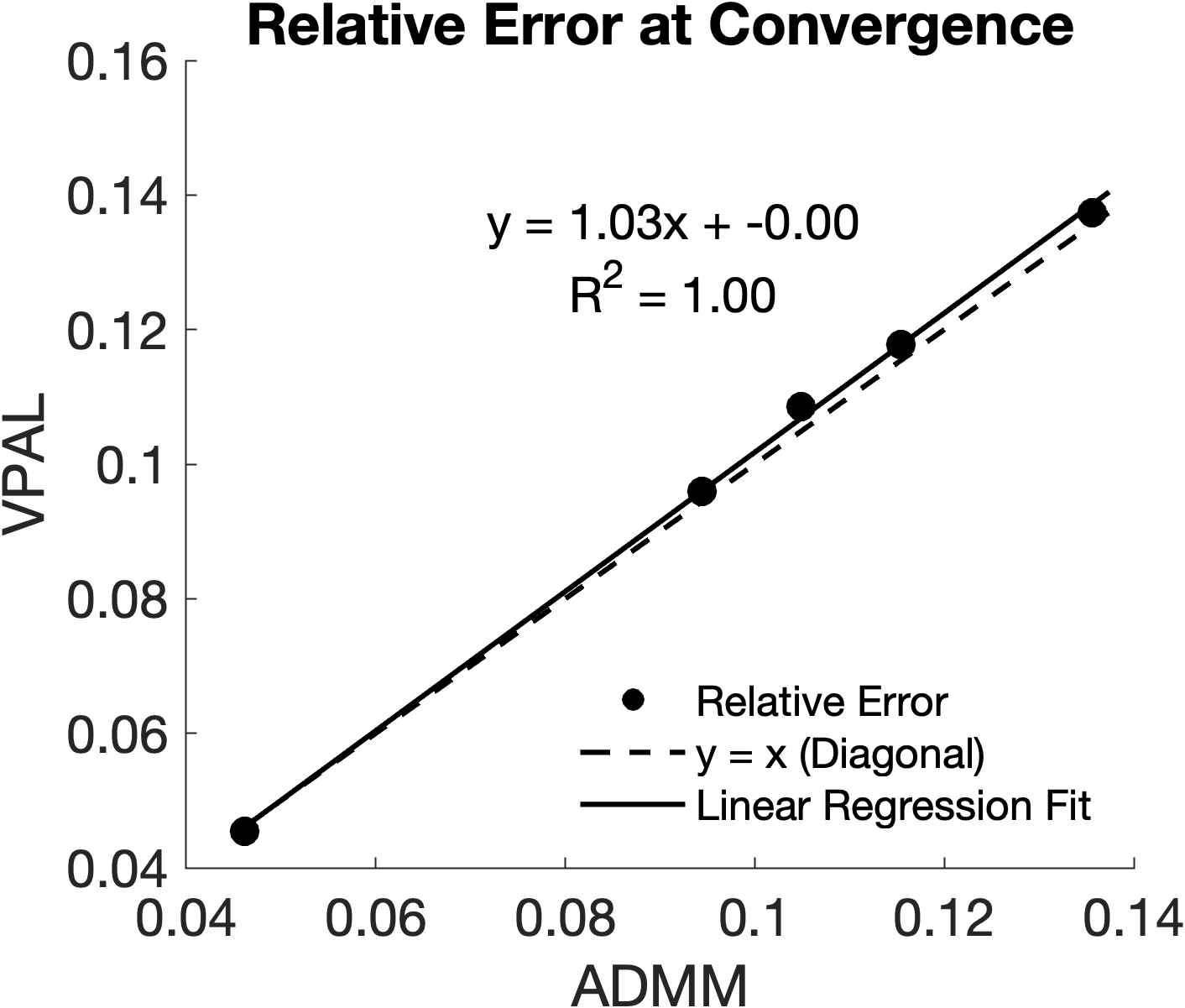}}
\caption{\textbf{Linear fitting of the relative error between ADMM and VPAL reconstructions with the ground-truth}  }\label{fig1}
\end{figure}

\subsection{In-vivo reconstructions}\label{subsec2}

In the 15 pediatric subjects, there are a varying number of cardiac frames, 17+/4 (mean and standard deviation), and the 4 (same for all subjects) respiratory frames to be reconstructed in the 5D free-running framework. The reformatted views of three example in-vivo subjects' ADMM and VPAL reconstructions are shown in Figure 4. 
\begin{figure}[t]
\centerline{\includegraphics[width=30pc,height=15pc]{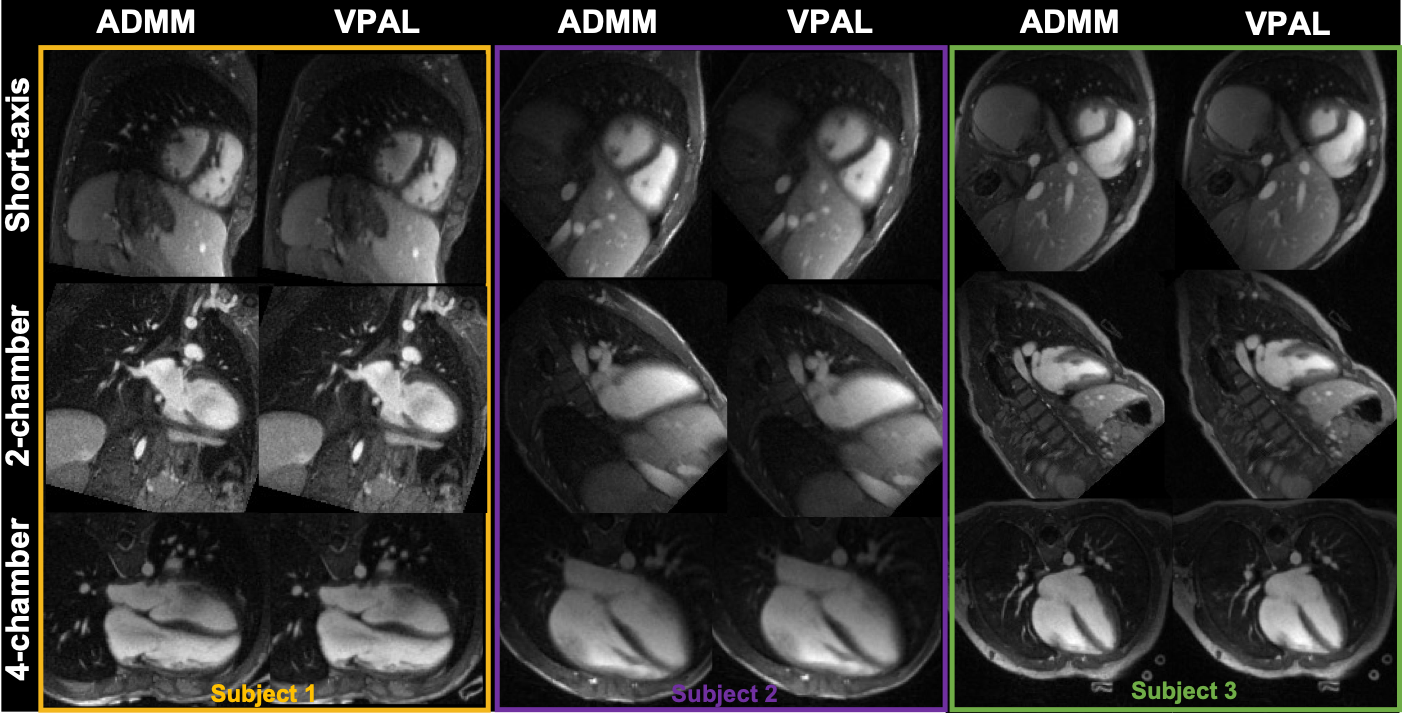}}
\caption{\textbf{Exampe in-vivo reconstructions using ADMM and VPAL on three subjects} The mid-short-axis, 2-chamber, and 4-chamber view in cardiac resting phase demonstrated in the output of the reconstructions using ADMM and VPAL algorithms.}\label{fig1}
\end{figure}

Using the same number of iterations as ADMM in reconstruction, VPAL is much faster to compute than ADMM, 4.7 $\pm$ 1.1 hours versus 16.3 $\pm$ 3.6 hours ($p=1e-10$). The reconstruction times of ADMM and VPAL are plotted in Figure 5. 
\begin{figure}[t]
\centerline{\includegraphics[width=20pc,height=15pc]{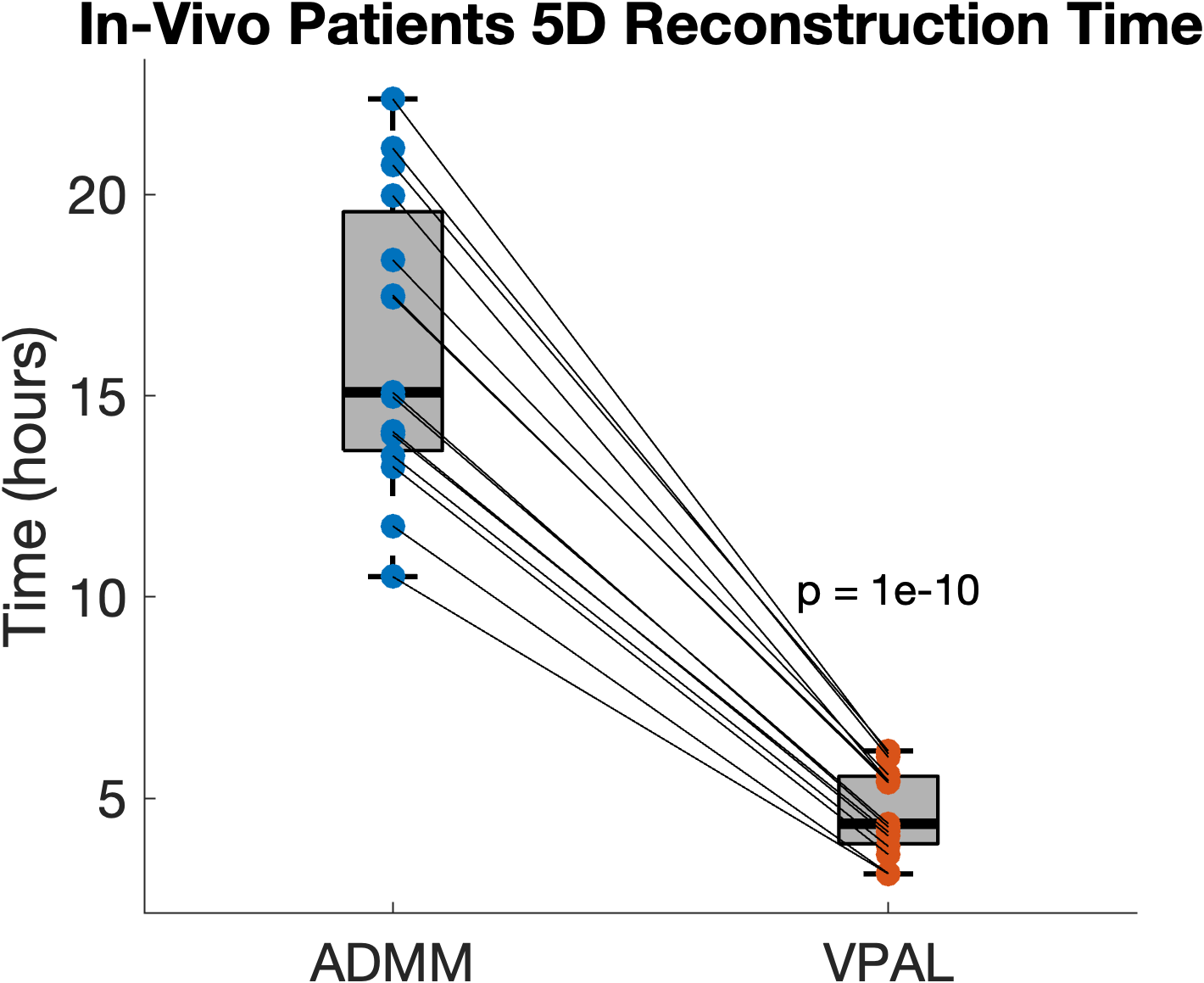}}
\caption{\textbf{Boxplot of reconstruction time using ADMM and VPAL algorithms.} On 15 in-vivo datasets, VPAL reconstruction is much faster than ADMM reconstruction.}\label{fig1}
\end{figure}
The mid-SA sharpness of VPAL is highly similar to that of ADMM, 0.076 $\pm$ 0.02 versus 0.075 $\pm$ 0.03, respectively. There is a strong linear relationship between the mid-SA sharpness of VPAL and ADMM with $R^2 = 0.97$, shown in Figure 6.
\begin{figure}[t]
\centerline{\includegraphics[width=20pc,height=15pc]{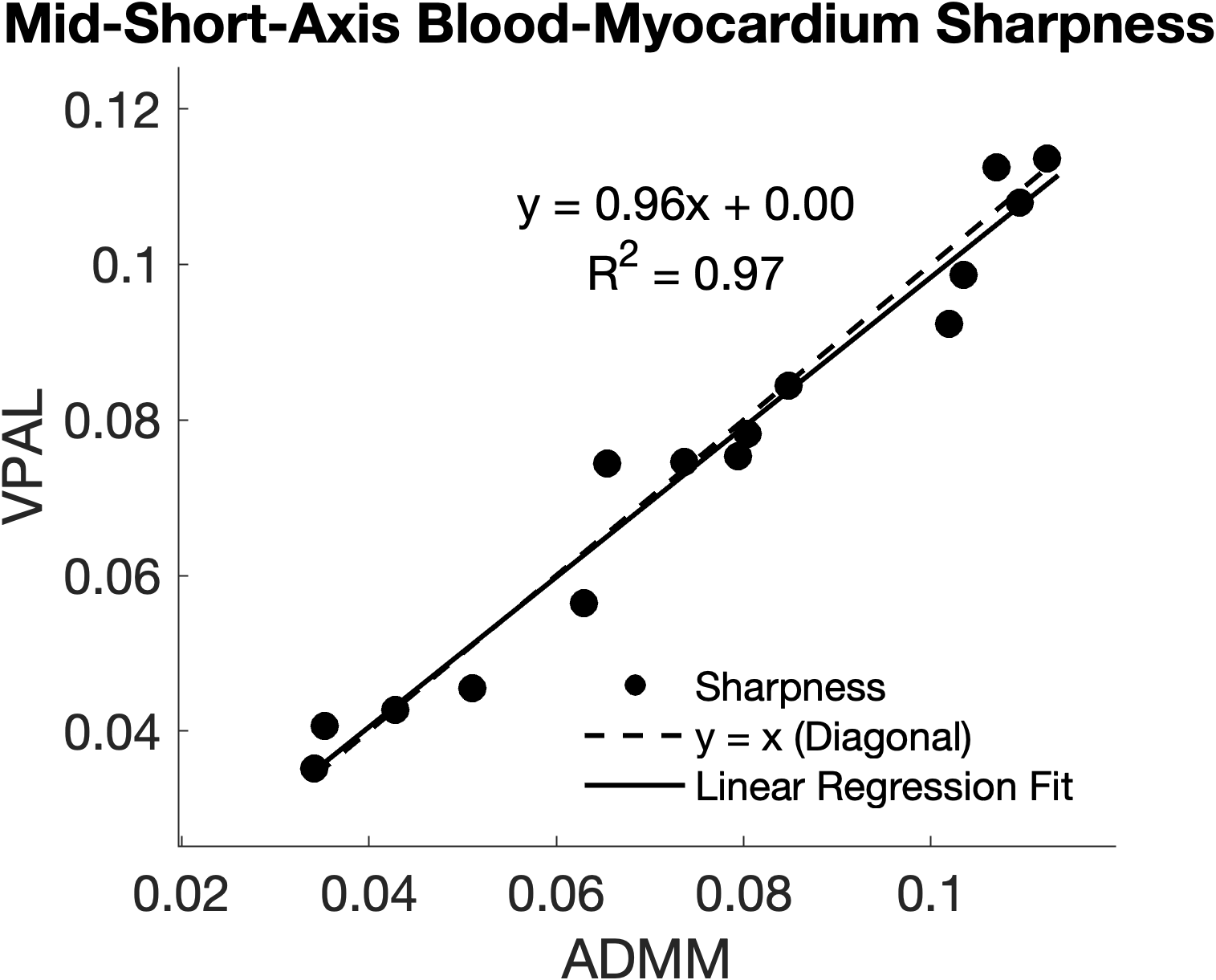}}
\caption{\textbf{Mid-SA sharpness of ADMM and VPAL reconstructions} On 15 in-vivo datasets, the mid-SA sharpness on VPAL reconstructions has a strong linear relationship with the ADMM reconstructions. $R^2 = 0.97$}\label{fig1}
\end{figure}

The difference between LVEF measured using VPAL versus ADMM reconstructions was not significantly different, 56 $\pm$ 6 \% versus 56 $\pm$ 6\%, respectively ($p=0.55$). The bland-altman plot in Figure 7 shows the 95\% confidence interval of agreement between the LVEF measurements using ADMM and VPAL reconstructions.  
\begin{figure}[t]
\centerline{\includegraphics[width=20pc,height=15pc]{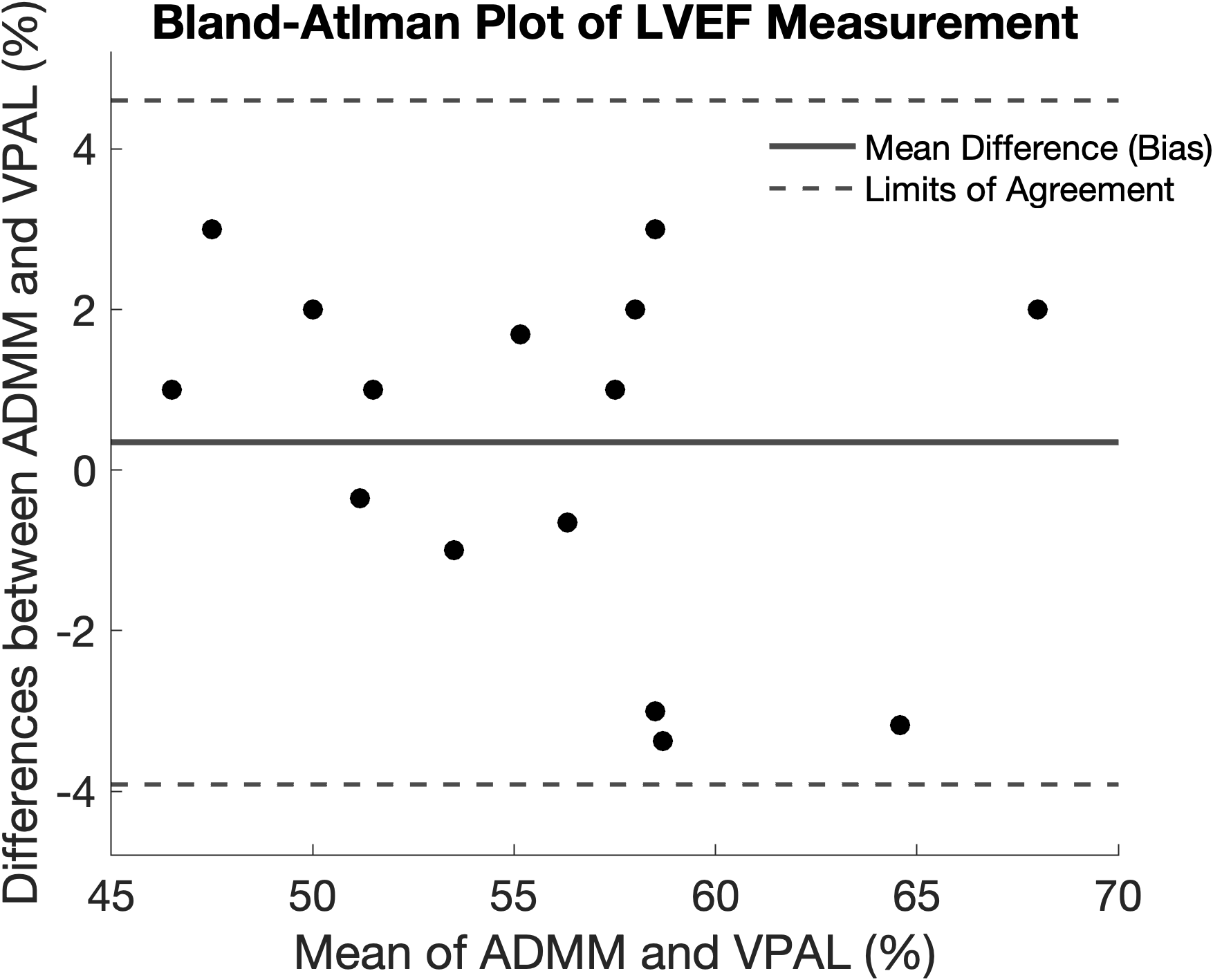}}
\caption{\textbf{Bland-Altman Analysis of LVEF of ADMM and VPAL reconstructions} On 15 in-vivo datasets, the difference between the LVEF measurements on VPAL and ADMM reconstructions is not significant using the 95\% confidence interval.}\label{fig1}
\end{figure}

The radiologist graded all cardiac cines in 2-chamber, 4-chamber and mid-short-axis cines to have good to excellent diagnostic values (3-4). The distribution of the image quality scores on ADMM and VPAL cines is shown in Figure 8. In 2-chamber reformatted cardiac cine images, ADMM images scored 3.8 $\pm$ 0.4 and VPAL scored 3.9 $\pm$ 0.3 (p=0.16). In 4-chamber reformatted images, ADMM scored 3.9 $\pm$ 0.4 and VPAL scored 3.9 $\pm$ 0.4 (p = 1), and in mid-short-axis images, ADMM scored 3.7 $\pm$ 0.5 and VPAL scored 3.7 $\pm$ 0.5 (p = 0.33). No significant differences were found in the radiologist's image quality ratings between ADMM and VPAL. 
\begin{figure}[t]
\centerline{\includegraphics[width=20pc,height=15pc]{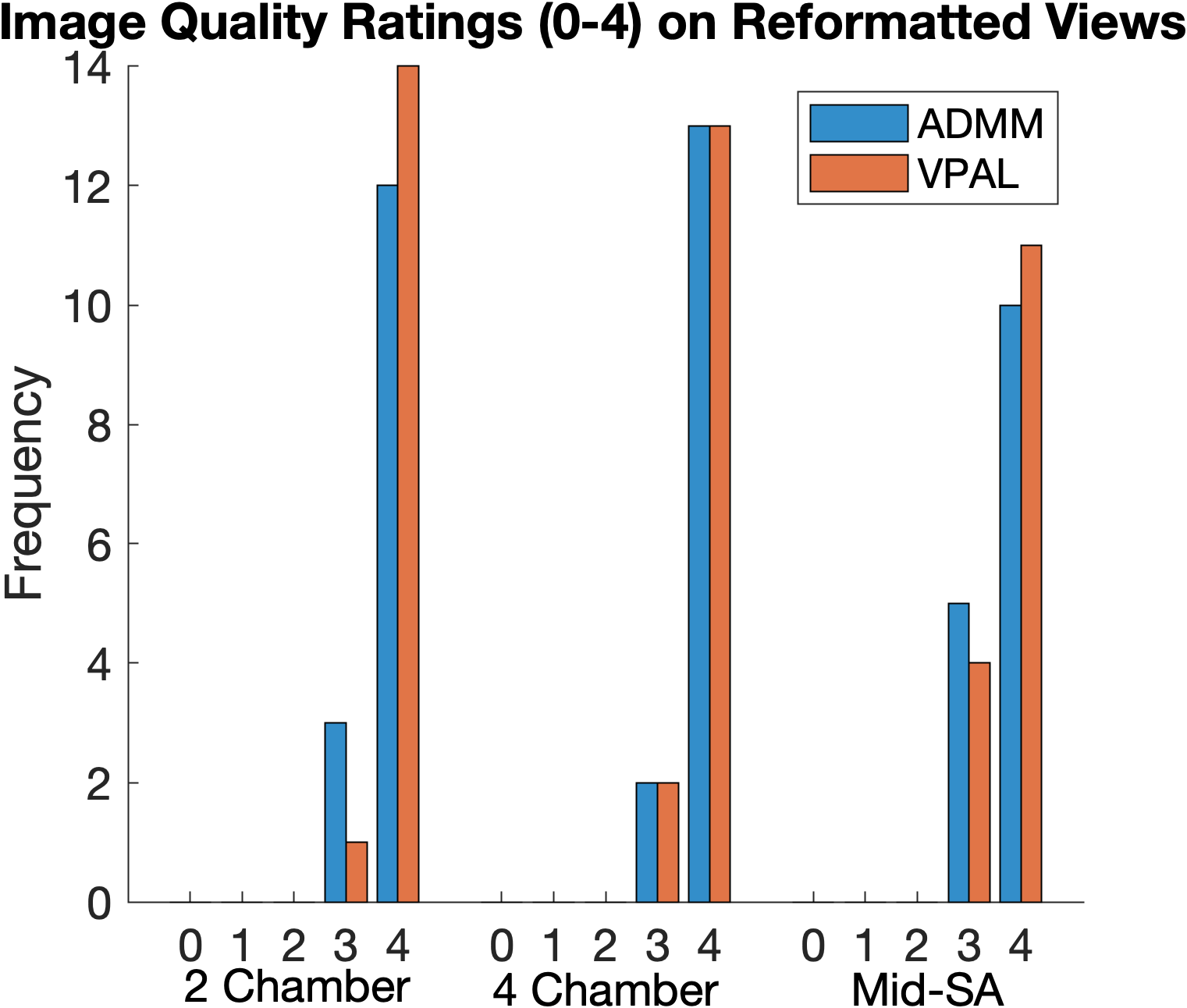}}
\caption{\textbf{Histogram of image quality ratings on cardiac cines of ADMM and VPAL reconstructions} On 15 in-vivo datasets, image quality scores are good (3) to excellent diagnostic (4) values.}\label{fig1}
\end{figure}
The 3D SSIM in the cardiac resting phase of the VPAL reconstructions is 0.98 $\pm$ 0.00, and in the peak systolic phase, it is also 0.98 $\pm$ 0.00.

\section{Discussion}
In this study, we introduced a variable projection augmented lagrangian (VPAL) method with a targeted update step using a Fletcher-Reeves nonlinear conjugate gradient to reconstruct the 5D cardiac- and respiratory-resolved images. Unlike the current state-of-the-art alternating-direction multiplier (ADMM) method that solves a full quadratic optimization problem in each iteration, VPAL updates using a single descent step on the variable projected problem, making it more computationally efficient. The main findings of this study were: 1) in numerical simulations, the convergence time is reduced by 27\% on average with the proposed VPAL algorithm comparing to the ADMM algorithm, although not statistically significant due to the large variance within subjects.  2) the relative error between the VPAL images at convergence and the ground truth images is strongly similar to that of the ADMM images and the ground truth images. 3) in in-vivo dataset, using the same number of iterations, the reconstruction time is greatly reduced (by 70\%) using VPAL compared to using ADMM.  4) The reconstructed VPAL images have the same image quality as the reconstructed ADMM images measured by mid-SA blood-myocardium border sharpness, 3D SSIM, and image quality scores. 5) The reconstructed VPAL and ADMM images have good agreement on the functional metric evaluated by left ventricular ejection fraction (LVEF). 

5D free-running imaging is a free-breathing technique that eliminates the unpredictable scan-time problem in navigator-gated 3D motion-resolved imaging and inaccurate gating in ECG-based acquisition. The high-resolution cardiac and respiratory resolved images of the entire heart enable precise delineation of the blood-myocardium borders, which are critical for accurate clinical evaluation of functional metrics such as ejection fraction and wall motion. However, the ill-posed reconstruction problem in 5D free-running imaging presents significant challenges due to the large scaleness of the problem, resulting in intensive memory usage and prolonged computation time. Using the current state-of-the-art reconstruction algorithm, such as ADMM, 5D free-running imaging technique is difficult to implement clinically due to ADMM's computational intensity. Instead, we demonstrate that VPAL, a novel numerical reconstruction approach, can effectively replace the ADMM method to reconstruct images from the undersampled k-space data. VPAL achieves reconstruction accuracy comparable to ADMM but with substantially reduced computation time. In in-vivo experiments, VPAL greatly shortens the reconstruction time while maintaining high image quality for functional evaluations such as ejection fraction, compared to ADMM. These results highlight VPAL as a more efficient reconstruction technique compared to ADMM, offer the advantage of clinical adaptation of a more streamlined and reliable workflow for advanced cardiac imaging, which alleviates the burden of breathholding, operator planning, and problems of unpredictable scan time, false gating in the current 3D motion-resolved imaging. 

An earlier study showed that VPAL converges faster than ADMM, mainly due to the elimination of the steps required to compute the exact or inexact variable $x$ in each iteration, which involves multiple gradient computations\cite{bib19}. In this study, we observed a similar trend on a large-scale problem using simulated dynamic 3D cardiac images. Specifically, the field of view (FOV) and matrix size of each numerical simulation were set identically to those in the in-vivo dataset. We found that VPAL converges much faster than ADMM, from 5.3 $\pm$ 2.4 hours (ADMM) to 3.9 $\pm$ 1.2 hours (VPAL). Although no significant differences in convergence time were found among the five numerical simulation datasets, likely due to low sample size and variability in noise distribution of the original functional CT images - the consistent reduction in convergence time across dataset aligns with previous finding on VPAL \cite{bib19}. This highlights the advantage of VPAL in accelerating high-dimensional reconstruction compared to ADMM, as computational time scales cubically with the number of voxels in each spatial dimension. In addition, we confirmed that the VPAL reconstructs the underlying ground-truth image with accuracy comparable to ADMM, as indicated by the close agreement in relative error between the VPAL and ADMM reconstructions when compared to the ground-truth image. 

VPAL can achieve the same results as ADMM with a shorter iteration time as it projects the solution of the variable $x$ in a set that minimizes the parts of the function related to the auxiliary variable $z$. However, each update step for $x$ is linearized in VPAL, which can have two potential problems 1) the linearization may be too simplified to encompass the complexity of the problem, potentially leading to more iteration steps to reach a solution and adversely affecting the computational efficiency, 2) for an ill-posed inverse problem such as MRI reconstruction, the linearization can introduce instability in the updates where small errors in the linear approximation maybe amplified, leading to numerical instability and unreliable results. In this work, We address these issues by introducing a non-linear Fletcher-Reeves update. The conjugate gradient update step ensures that each update step is orthogonal to the previous step, which could improve the convergence speed. The nonlinear update plays a stronger role as it captures the nonlinearity in the problem, $\ell1$ regularization, and potentially replaces multiple CG iterations to solve $x$ by remembering previous steps. The new CG step is a linear combination of the new search direction and the weighted CG step of the previous iteration.

In the study, We implemented one variation of the nonlinear conjugate gradient method, the Flecher-Reeves (FR) method, within the VPAL reconstruction framework. The FR method is the commonly used approach for updating search directions in non-linear CG algorithms. Other nonlinear conjugate gradient algorithms, such as Polak-Ribiere (PR) or Hestenes-Stiefel (HS), could potentially offer better convergence rates or robustness. However, the full exploration of other nonlinear CG methods is not the primary aim of this study; instead, we mainly aim to demonstrate that VPAL provides a computationally efficient alternative to the ADMM reconstruction algorithm. By reducing computational complexity, VPAL has the potential to facilitate the clinical adoption of 5D free-running imaging.

In this study, we implemented the linear VPAL step size update within the line search process. This approach assumes that the variable $z$  is already optimized in the new update step and therefore does not contribute to the search direction during the line search. Although this simplification is computationally efficient, previous studies have shown that using a non-linear VPAL step size calculation can improve performance. Future studies will explore the integration of nonlinear step size updates to further enhance the effectiveness of the algorithm.

One of the key advantages of using VPAL in the underdetermined, ill-posed, and large-scale image reconstruction problem is that each iteration is much faster than the ADMM iterations, which gives VPAL a unique opportunity for systemic optimization of regularization parameters. The regularization parameters in the reconstruction process are often empirically selected based on visual assessments of the reconstructed images to balance the trade-off between data fidelity, which increases the noise level, and regularization, which strengthens the sparsity assumption and increases smoothness in the image. The systemic optimization of regularization parameters that have been explored previously include 1) using neural networks as surrogates for optimization that extract features from input data (e.g. noise level, contrast) \cite{bib14} and use supervised learning to map these features to optimal parameters \cite{bib32}, 2) using reinforcement learning for adaptive tuning during the iterative VPAL reconstruction process \cite{bib33}, and 3) using a hybrid approach that uses a neural network to initialize the regularization parameters and refines the parameters during VPAL iterations using optimization methods such as line search or Bayesian optimization \cite{bib34}. The adaptation of VPAL for systemic optimization of regularization parameters could potentially improve the convergence speed of VPAL and therefore accelerate the reconstruction of 5D free-running imaging, which will be explored in future studies. 

In this study, our formulation of the reconstruction problem employs $\ell_2$-norm regularization on the temporal variations, which encourages smooth cardiac and respiratory motion. Although this formulation differs from the standard formulations \cite{bib6} that employs $\ell_1$-norm regularization on the temporal finite difference, the purpose of this study is to compare the efficiency of ADMM and VPAL optimizers.  Therefore, the comparison between the proposed algorithm VPAL and ADMM remains valid because both methods are general optimization frameworks that can be applied to a wide range of regularization functions, including both $\ell_1$ and $\ell_2$ norms. Our findings regarding computational efficiency, convergence behavior, and stability of VPAL approach can still provide valuable insights for solving the $\ell_1$-norm-regularized problem. Future work may extend this comparison to the $\ell1$ norm formulation to further explore the applicability of VPAL in enforcing sparsity in temporal dynamics while maintaining computational efficiency. 

This study has several limitations: 1) this study is limited by the availability of datasets, which restricts the breadth of evaluation; 2) the k-space data is retrospectively simulated by directly using the forward operator on the 5D image data derived from the 4D functional CT image without introducing additional noise or blurring, which restricts the reconstruction performance comparison to one signal-to-noise setting; 3) coil sensitivity encoding is not enforced in the numerical simulation, which is not consistent with the in-vivo reconstruction scenario as it results in a more under-determined problem.

\section{Conclusion}
We introduce VPAL as a more efficient algorithm for iterative reconstruction of 5D free-running whole heart data compared to the state-of-the-art ADMM methods. Using the same number of iterations, the VPAL algorithm reconstructs images significantly faster than the ADMM algorithm while maintaining the same image quality, as evaluated by mid-short axis sharpness of the blood myocardium, 3D SSIM, image quality ratings from the radiologists, and functional quantification using LVEF.  VPAL has the potential to be utilized for efficient regularization parameter tuning and streamline the use of 5D free-running motion-resolved imaging. 

\section{Acknowledgment}
This study is partially funded by the 2023--2024 Emory URC interdisciplinary award (Chung and Oshinski) and the National Institute of Biomedical Imaging and Bioengineering (NIBIB) R01-EB027774 (Oshinski).  This work was partially supported by the National Science Foundation (NSF) under grant DMS-2152661 (Chung)

\bibliography{sn-bibliography}% common bib file

\end{document}